\documentstyle[12pt,twoside,fleqn,espcrc1,epsfig,glas,subfigure]{article}
\newlength{\dinwidth}
\newlength{\dinmargin}
\newcommand{\mean}[1]{\left< #1 \right>}
\newcommand{\fmean}{\mean{F}}
\newcommand{\fpert}{\fmean^{\rm pert}}
\newcommand{\fpow}{\fmean^{\rm pow}}
\def\st{\scriptstyle}

\def\as{\alpha_{s}}

\def\abp{\overline{\alpha}_{p-1}}
\def\anotbar{\overline{\alpha}_{0}}
\def\be{\begin{equation}}
\def\ee{\end{equation}}
\def\bea{\begin{eqnarray}}
\def\eea{\end{eqnarray}}

\def\lsim{\mathrel{\rlap{\lower4pt\hbox{\hskip1pt$\sim$}}
    \raise1pt\hbox{$<$}}}                % less than or approx. symbol
\def\gsim{\mathrel{\rlap{\lower4pt\hbox{\hskip1pt$\sim$}}
    \raise1pt\hbox{$>$}}}                % greater than or approx. symbol
\begin{document}
\begin{titlepage}{GLAS-PPE/2000-05}{3$^{\underline{\rm{rd}}}$ August 2000}
\title{Event Shapes and Power Correction Results from HERA}
\author{G.J. McCance
%\address{Dept. of Physics and Astronomy,
%        University of Glasgow, \\
%        Glasgow G12 8QQ, United Kingdom}
}
\vspace{0.3cm}
\centerline{\em Presented at QCD and High Energy Hadronic Interactions}
\centerline{\em Rencontres de Moriond 2000}
\centerline{\em On behalf of the H1 and ZEUS collaborations}
\vspace{0.8cm}
\begin{abstract}
%\noindent
Inclusive event shape variables have been measured in the Breit Frame for neutral current deep-inelastic positron-proton scattering using the H1 and ZEUS detectors at HERA. The variables thrust, jet broadening, $C$-parameter, jet mass and two kinds of differential two-jet rate have been studied in the kinematic range $7 < Q < 100$ $GeV$. The $Q$ dependence of the shape variables have been compared with QCD applying power corrections proportional to $1/Q^{p}$ to account for hadronisation effects. The concept of power corrections is tested by fitting the strong coupling constant $\as$ and a non-perturbative parameter $\abp$.
\end{abstract}
\end{titlepage}
\section{Introduction}
A recent revival of interest in event shapes has been prompted by theoretical developments in the understanding of hadronisation \cite{paper:power}. These approaches extend perturbative QCD calculations into the region of low momentum transfers using approximations to higher-order graphs; the hadronisation contributions are expected to go as $1/Q^{p}$ where $p$ is an integer power ($1$ or $2$). The $Q^2$ scale at HERA can be varied over four orders of magnitude allowing these corrections to be studied in detail. The data are compared to theoretical expectations for power corrections, characterised by $\as(M_Z)$ and an effective coupling $\abp(\mu_I)$ specified at the infrared matching scale $\mu_I~\simeq~2$~$GeV$, and Next-to-Leading Order (NLO) perturbative QCD (pQCD) calculations, determined by $\as(M_Z)$.

A suitable frame of reference in which to study event shapes is the Breit frame since this frame maximises the separation between the current jet from the struck quark and the proton remnant. Viewed in this frame the exchanged gauge boson is purely spacelike and the incoming QPM quark is back-scattered with equal and opposite momentum ($Q/2$). The Breit frame also permits the event shapes from the spacelike process in DIS to be compared directly to the event shapes in the timelike $e^+e^-$ process. Two classes of event shapes are studied.  

\begin{figure}[p]
  \centering
{\epsfig{figure=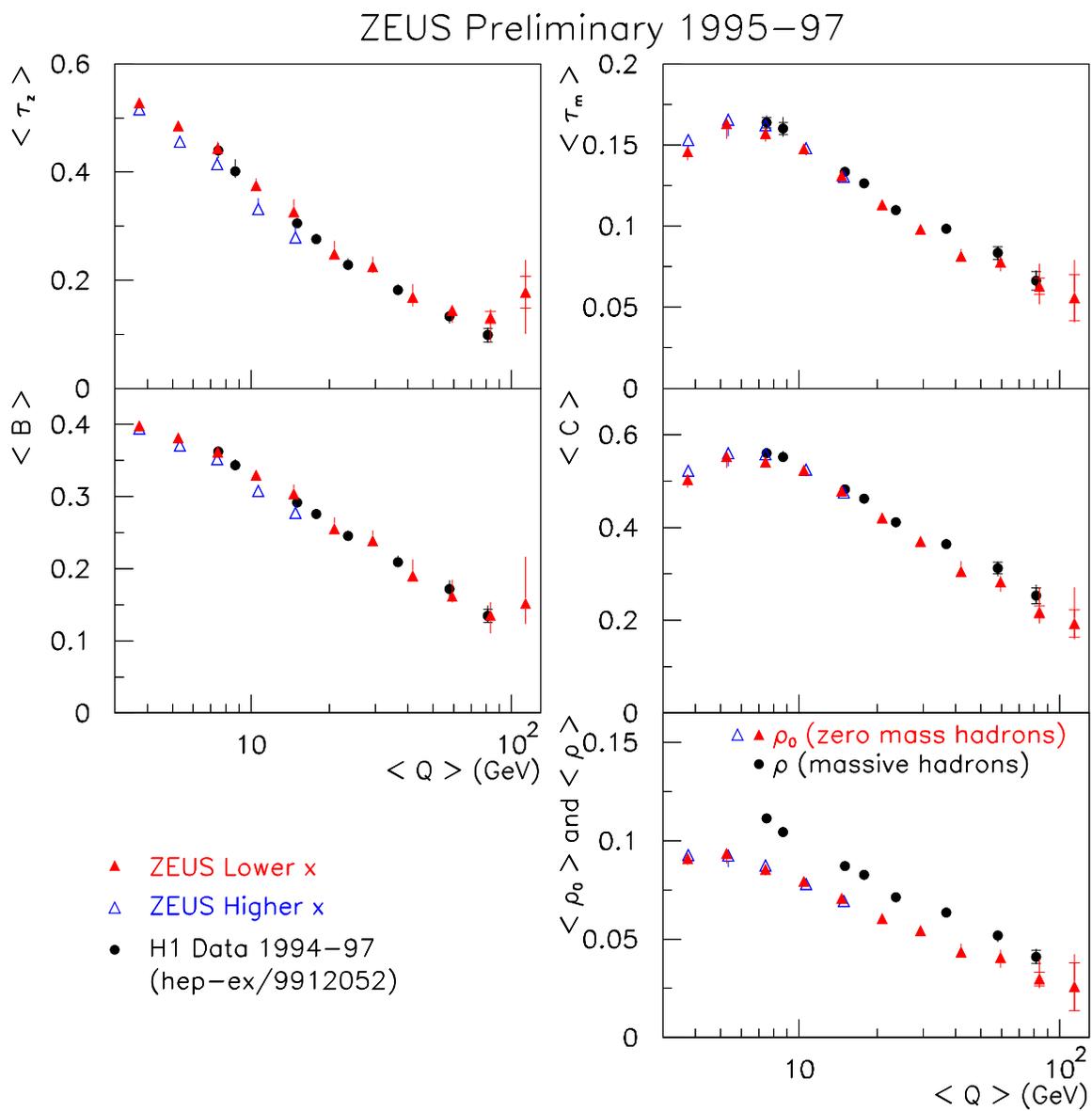,width=.9\textwidth}}\quad
  \caption[]{Comparison of H1 and ZEUS data. Note the $x$-dependence in the ZEUS data.}
\label{fig:zeush1}
\end{figure}

The first class are those confined to the current region of the Breit frame ($z<0$), namely, thrust, both with respect to the Breit frame axis ($\tau_z$) and the thrust axis ($\tau_{max}$), jet broadening with respect to the Breit frame axis ($B$), jet-mass ($\rho$), and the $C$-parameter (C).
%\begin{equation}
%\tau_z = 1 - \frac{\sum_{i}{|\vec{n}.\vec{p}_{i}|}}{\sum_{i}{|\vec{p}_{i}|}}
%    = 1 - \frac{\sum_{i}{p_z}}{\sum_{i}{|\vec{p}_{i}|}}    \label{eqn:tz}
%\end{equation}
%
%\begin{equation}
%\tau_{max} = 1 - \frac{\sum_{i}{|\vec{t}.\vec{p}_{i}|}}{\sum_{i}{|\vec{p}_{i}|}}
%    \label{eqn:tzmax}
%\end{equation}
%
%\begin{equation}
%B = \frac{\sum_{i}{|\vec{n}\times\vec{p}_{i}|}} {2 \sum_{i}{|\vec{p}_{i}|}}
%    = \frac{\sum_{i}{p_{\bot}}}{2 \sum_{i}{|\vec{p}_{i}|}}   \label{eqn:jbroad}
%\end{equation}
%
%\begin{equation}
%\rho = \frac{M^{2}}{{(2 E_{tot})}^{2}}
%       = \frac{ {(\sum_{i} p^{\mu}_{i})}^{2} } { 4{(\sum_{i} E_{i})}^{2} }    %\label{eqn:jmass}
%\end{equation}
%
%\begin{equation}
%C = \frac{3}{2(\sum{|p_i|})^{2}}\sum_{i,j}|p_{i}||p_{j}|\sin^{2}\theta_{ij}  \label{eqn:cpar2}
%\end{equation}
%Note the definition given for $C$-parameter in equation \ref{eqn:cpar2} is analytically equivalent to the more standard eigenvalue based definition \cite{paper:resummedc}. 
Thrust ($\tau_z$) and jet broadening ($B$) are measured with respect to the Breit frame axis and are therefore specific to DIS; the remainder can also be defined in $e^+e^-$ experiments. At the hadron level, the H1 result \cite{paper:h1} assumes the hadrons have their true mass, the ZEUS result assumes the hadrons have zero mass; this difference in definition only affects the jet-mass which can be seen in Fig. \ref{fig:zeush1}. Here the data are compared and the other variables show agreement between the two data sets. ZEUS bin in kinematic ranges of both $x$ and $Q$, so in some regions ZEUS have two data points measured over the same range in $Q$ but a different range in $x$; H1 integrate over $x$. In order to keep the variables infrared safe, the total energy in the current region has to exceed 10\% (3\%) of the total available energy, for H1 (ZEUS). The exact value of this cut is not critical.
The second class of variables are the two-jet rates. These are measured over the full phase space of the Breit frame in which $(n+1)$ jets are searched for; the $+1$ denotes the proton remnant. Both the $k_t$ (or Durham) and the factorisable JADE scheme are used in which a jet merging criteria based on distance measures $y_{ij}$ between pairs of particles is defined. The event shape $y$ denotes the value of $y_{ij}$ at which the transition $(2+1) \rightarrow (1+1)$ jets occurs.

The mean values of these event shapes can be interpreted theoretically as
\begin{equation}
       \fmean =\fpert + \fpow                                                  \label{eqn:sumit}
\end{equation}
where
\begin{equation}
  \fpert = c_{1,F}(x,Q)\cdot\as(Q) + c_{2,F}(x,Q)\cdot\as^2(Q)                   \label{eqn:c1c2}
\end{equation}
with $c_{1,F}(x,Q)$ and $c_{2,F}(x,Q)$ being calculable coefficients of pQCD, here determined by the DISENT NLO program \cite{paper:disent}
with renormalisation/factorisation scales set at $\mu_R=\mu_F=Q$.
A simple form for $\fpow$ of $A/Q$ or $B/Q^{2}$ has been ruled out by H1 data \cite{paper:h1}. A more sopisticated approach \cite{paper:power} introduces
\begin{equation}
\fpow = {\cal M}^\prime a_{F} \frac{16}{3 \pi p}  \left({\frac{\mu_I}{\mu_R}}\right)^{p}
  \bigg[ \abp(\mu_I) - \as(\mu_R) - \frac{23}{6\pi} \left( \ln\frac{\mu_R}{\mu_I} + 0.45 + \frac{1}{p} \right) \as^2(\mu_R) \bigg]   \label{eqn:powc}
\end{equation}
where $p$ is the power, $a_F$ is a calculable $F$-dependent coefficient and ${\cal M}^\prime = 2{\cal M}/\pi \simeq 0.95$ \cite{paper:newmilan} is a two-loop level refinement to the calculations called the Milan factor.

\section{Results}
\begin{figure}[p]
  \centering
{\epsfig{figure=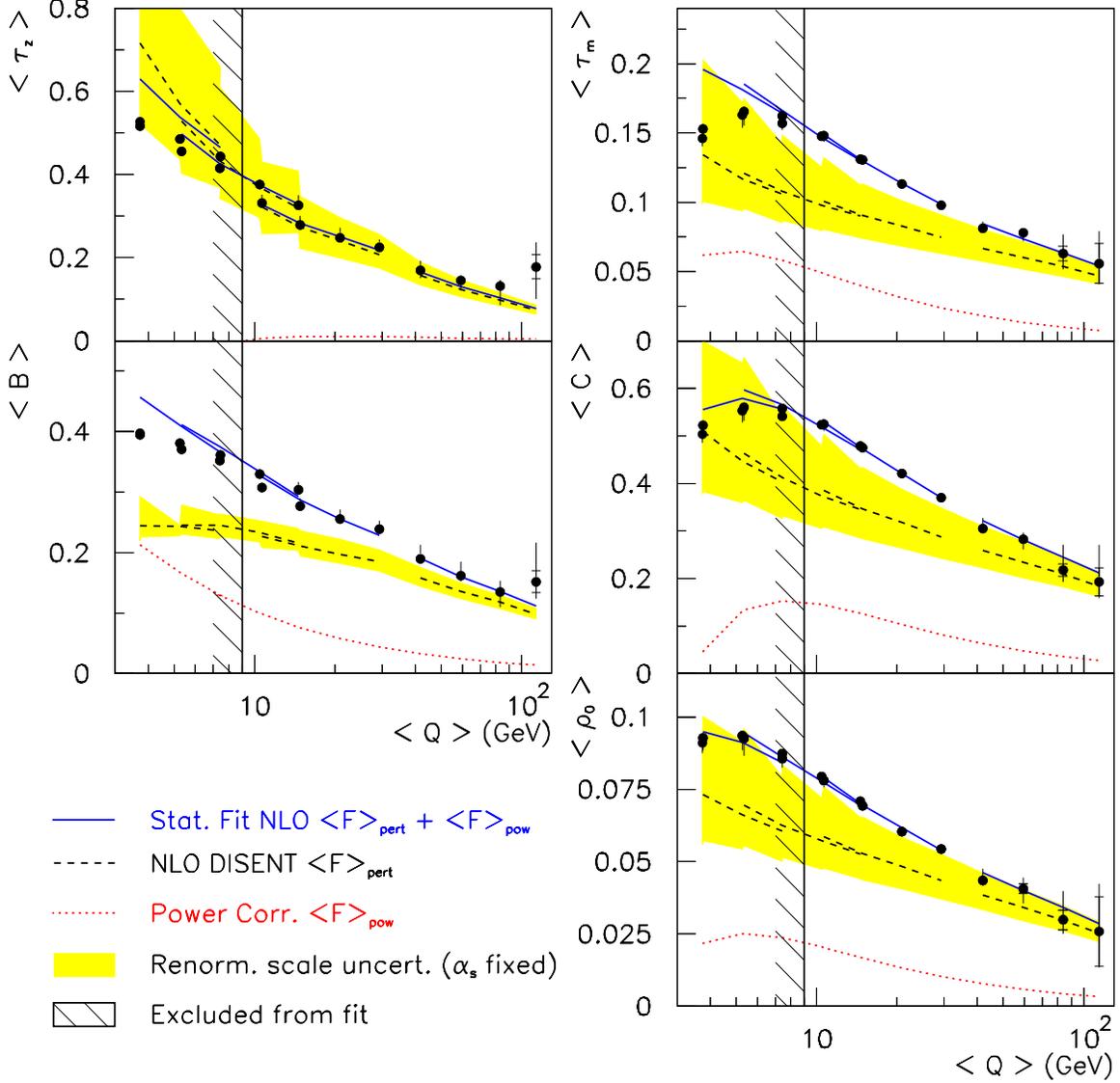,width=.9\textwidth}}
  \caption[]{Power correction fit to ZEUS data. The dashed line is the NLO $\fpert$ contribution, with the shaded band showing the effect of varying the renormalisation scale by a factor 2. The dotted line is the $\fpow$ power contribution and the full line is the sum of the two. The region below $9$~$GeV$ is excluded from the fit.}
\label{fig:zeusfit1}
\end{figure}

% Not sure why LaTeX screws up on figures 3 and 4 --> put in explicitly.
Fits of Eq (\ref{eqn:sumit}) to ZEUS data are plotted in Fig. \ref{fig:zeusfit1}. The shaded band indicates the theoretical uncertainty on $\fpert$ from varying the renormalisation scale by a factor of $2$ to $\mu_R=2\cdot Q$ and $\mu_R=1/2\cdot Q$; this is the dominant uncertainty on the measurement. The fitted values of the two parameters $\as(M_Z)$ and $\anotbar(\mu_I)$ from H1 and ZEUS data are plotted in Figs. 3 and 4 as contours (theoretical errors such as the scale dependence are not included in the contours). The H1 result shows the $\chi^2 = \chi^2_{min} + 1$ and $\chi^2 = \chi^2_{min} + 4$ contours for statistical and experimental systematic errors combined in quadrature. The smaller contours in the ZEUS result are a statistical only fit, the larger ones are the $\chi^2 = \chi^2_{min} + 4$ contours for the statistical and systematic errors combined in quadrature. Large correlations between $\as(M_Z)$ and $\anotbar(\mu_I)$ are seen for $\mean{\tau_z}$ but less for the other variables. Both experiments fit an approximately universal $\anotbar(\mu_I) \simeq 0.5 \pm 0.1\%$. The fitted values of $\as(M_Z)$ for the $e^+e^-$-type variables ($C$-parameter, jet-mass and thrust w.r.t. thrust axis) are compatible for the ZEUS result and remarkably close for the H1 result. Including the DIS specific variables (thrust and broadening w.r.t. Breit axis) sees a larger spread in the fitted values of $\as(M_Z)$. Possible explanations are missing higher order QCD corrections or incomplete knowledge of the power correction coefficients.
\begin{figure}[p]
\begin{center}
\epsfig{figure=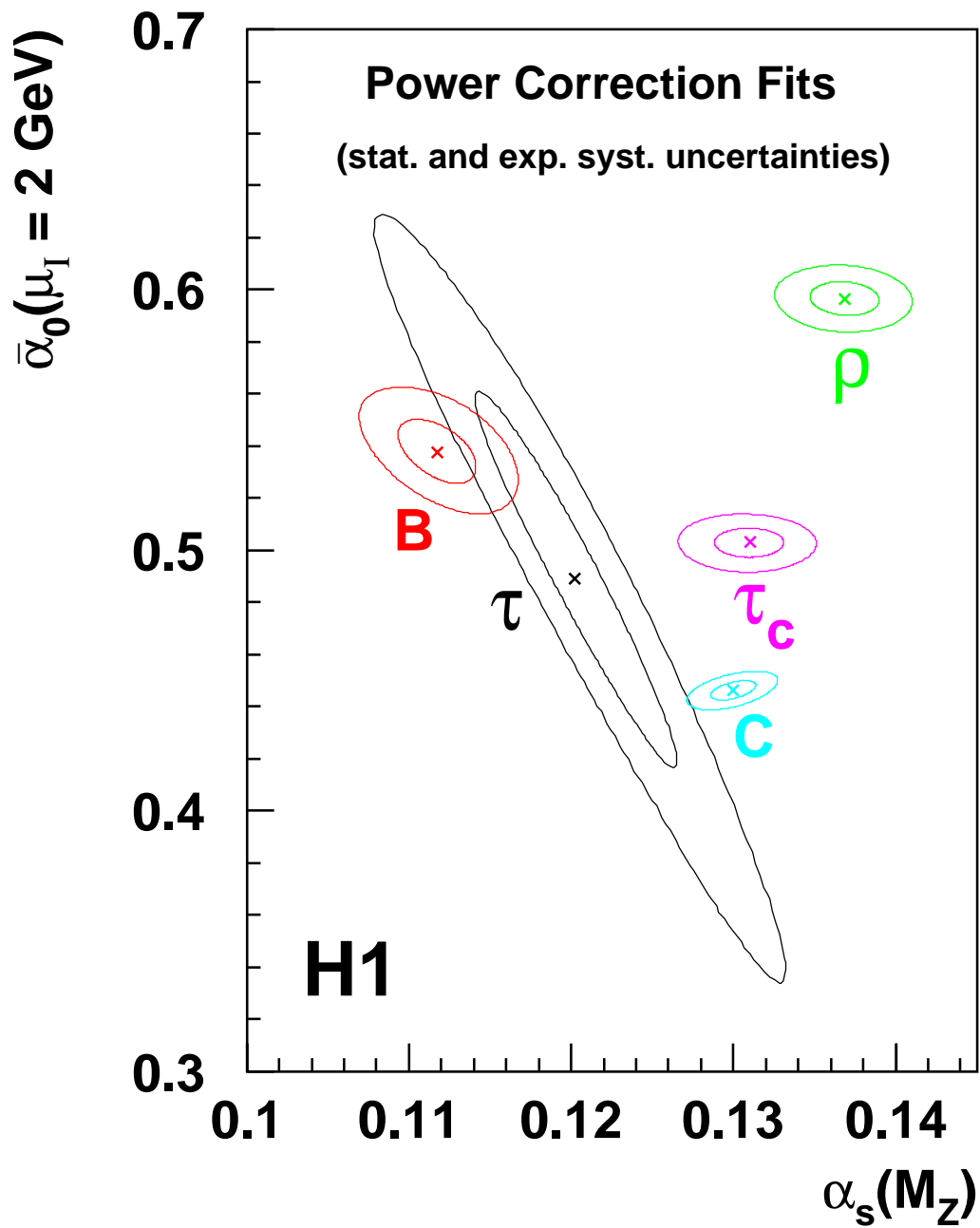,width=0.8\textwidth}
\caption{H1 contours showing fitted values of $\as(M_Z)$ and $\anotbar(\mu_I)$. The plot shows the statistical $\oplus$ systematic contours for two confidence levels for each of the variables.}
\end{center}
\label{fig:contourh1}
\end{figure}
\begin{figure}[p]
\begin{center}
\epsfig{figure=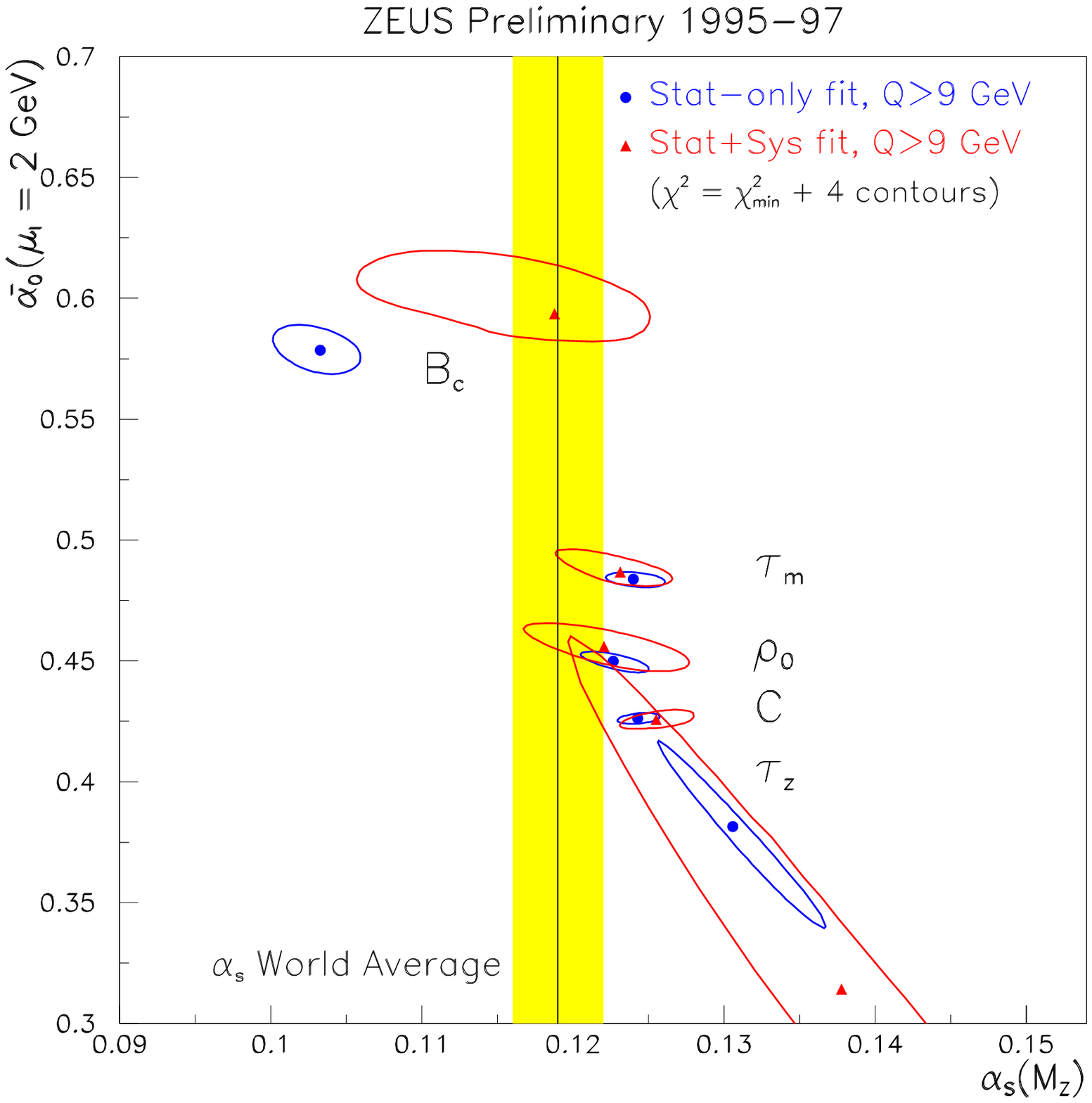,width=0.9\textwidth}
\caption{ZEUS contours showing fitted values of $\as(M_Z)$ and $\anotbar(\mu_I)$. The plot shows both the statistical only contours and the statistical $\oplus$ systematic contours for each of the variables.}
\end{center}
\label{fig:contourzeus}
\end{figure}
Fits to H1 two-jet rate data are plotted in Fig. \ref{fig:h1jet} \cite{paper:h1}. The hadronisation correction is seen to be small for these variables. For the $\mean{y}$ from the factorisable JADE algorithm, table \ref{table:yfj} shows that the fitted value of $\anotbar(\mu_I)$ is unreasonably low with the conjectured $a_F$ factor of $1.0$. Instead, a much more reasonable fit can be obtained by allowing the power corrections to become small and negative, with an $a_F$ factor of $-0.2$.
\begin{table}[h]
  \centering
  \caption{Fits of $\mean{y}$ to H1 data showing preference for negative power corrections.}
  \begin{tabular}{|c|c|c||c|c|c|}
    \hline
    \multicolumn{1}{|c|}{$\fmean$} &
    \multicolumn{1}{|c|}{$a_F$} &
    \multicolumn{1}{|c||}{$p$} &
    \multicolumn{1}{|c|}{$\abp(\mu_{I}=2$ $GeV)$} &
    \multicolumn{1}{|c|}{$\as(M_Z)$} &
    \multicolumn{1}{|c|}{$\chi^2/{dof}$} \\\hline
    $\mean{y_{fJ}}$ & $1$ & $1$ &
    $0.28~^{+0.02}_{-0.02}$ &
    $0.105~^{+0.005}_{-0.006}$ &
    $0.8$ \\\hline
    $\mean{y_{fJ}}$ & $-0.2^*$ & $1$ &
    $0.37~^{+0.20}_{-0.21}$ &
    $0.116~^{+0.008}_{-0.009}$ &
    $0.6$ \\\hline
  \end{tabular}
    \label{table:yfj}
\end{table}

The power fit for the $\mean{y}$ from the $k_t$ algorithm is of special interest because the power $p$ is expected to be 2 instead of 1 as it is for all the other variables. The H1 data are inconsistent with a fit using $p=1$. The data are consistent with $p=2$ but a more qualitative statement cannot be made given current experimental precision on this variable and the fact that the theoretical $a_F$ factor for this variable is unknown.

\begin{figure}[p]
\begin{center}
\epsfig{figure=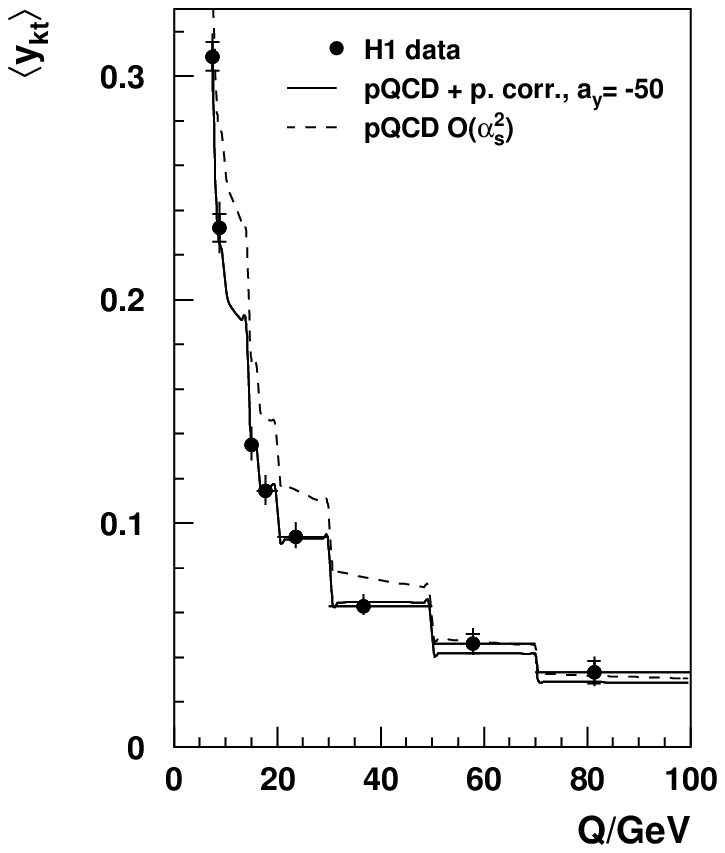,width=0.48\textwidth}
\hspace{0.3cm}
\epsfig{figure=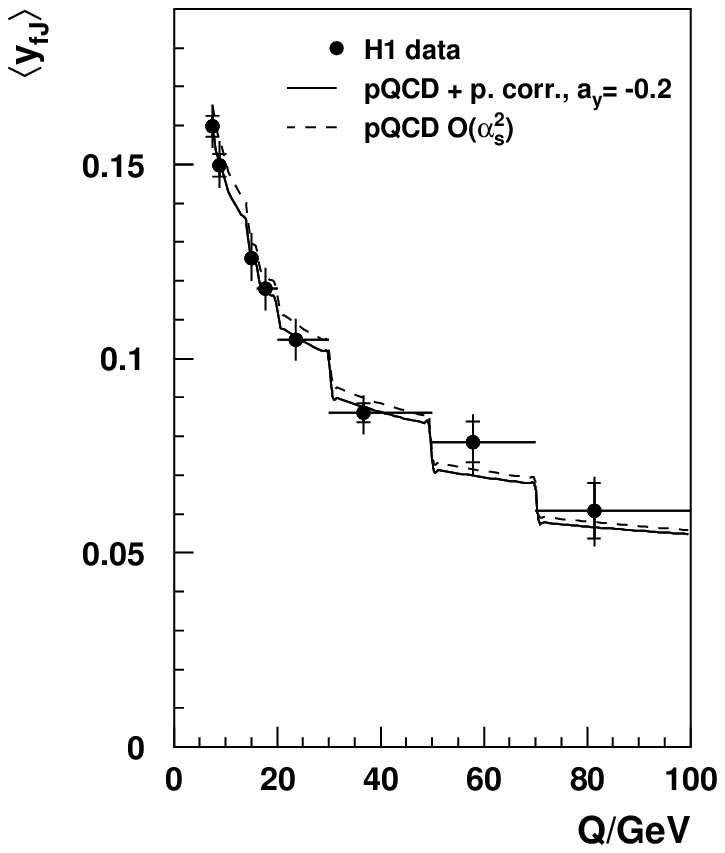,width=0.48\textwidth}
\end{center}
\caption{Fits of $\mean{y}$ H1 data, for $k_t$ (left) and factorisable JADE (right) algorithms. The dotted line is the perturbative QCD prediction only, the solid line is this prediction plus the power correction.
\label{fig:h1jet}}
\end{figure}

\begin{figure}[p]
\begin{center}
\epsfig{figure=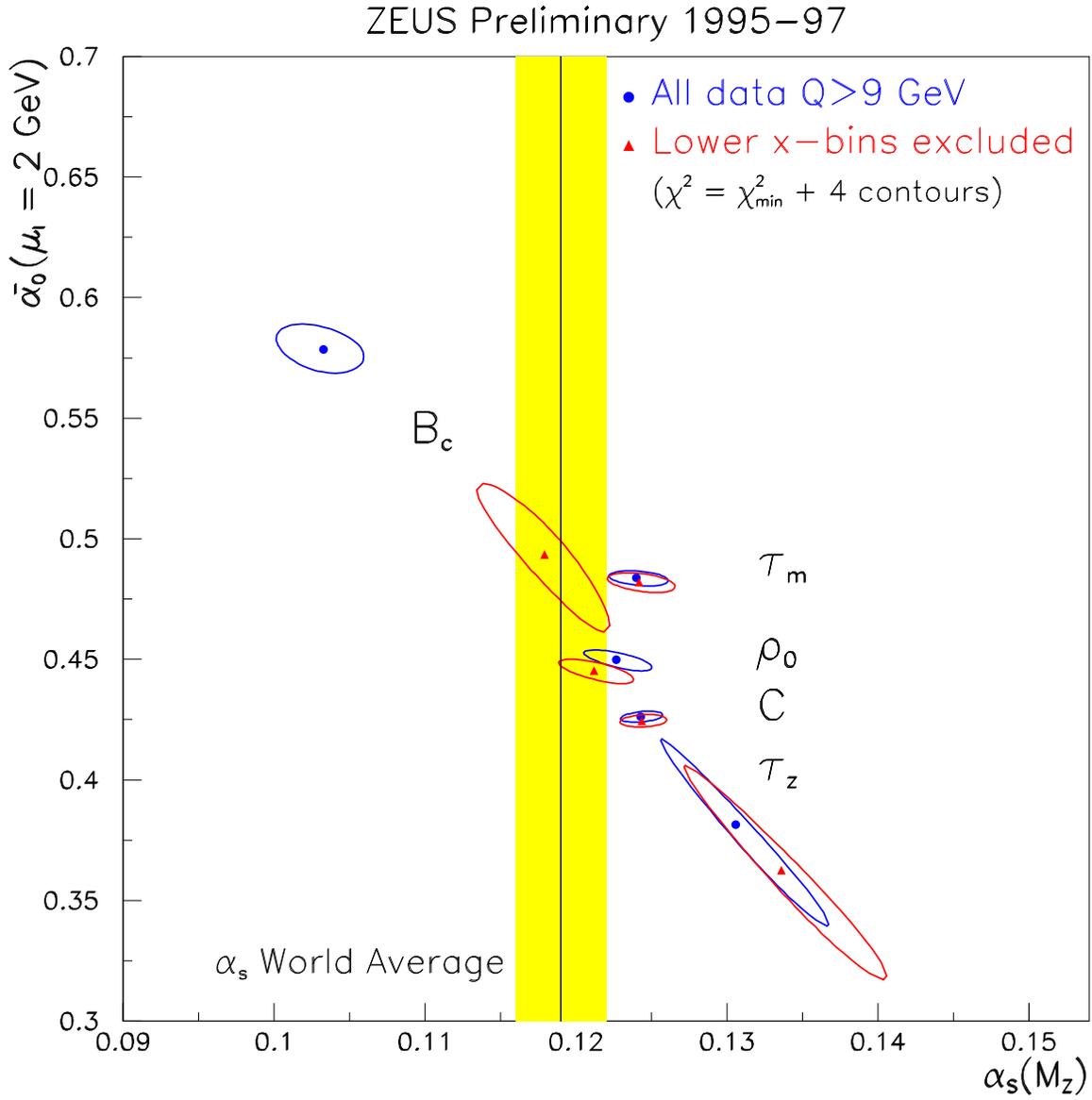, width=0.9\textwidth}
\caption{Fit to ZEUS data showing a clear shift in the contour for jet broadening $B_c$ due to the $x$-dependence. Plotted are the statistical only contours.
\label{fig:zeusx}}
\end{center}
\end{figure}

ZEUS have also studied the $x$-dependence of the power correction fits. Figure \ref{fig:zeusx} shows the result of omitting the lower $x$ bins from the ZEUS data given in Fig. \ref{fig:zeush1}; plotted are the statistical contours only. Most of the variables display little sensitivity to $x$. However, a large effect is seen in the jet broadening where the fitted values of $\as(M_Z)$ and $\anotbar(\mu_I)$ shift considerably. This confirms recent theoretical comment \cite{paper:revisit} that the $a_F$ factor for jet broadening should contain $x$-dependent terms that are, as yet, unknown.

In conclusion, this new theoretical approach to hadronisation works very well given its simplicity, although there remain some theoretical questions particularly in the $x$-dependence of the DIS specific variables.

%\section*{Acknowledgments}
%This is where one places acknowledgments for funding bodies etc.
%Note that there are no section numbers for the Acknowledgments, Appendix
%or References.


\begin{thebibliography}{99}
\bibitem{paper:power}Yu.L. Dokshitzer and B.R. Webber, {\em Phys. Lett.} {\bf B 404} (1997) 321.\\
Yu.L. Dokshitzer, A. Lucenti, G.Marchesini, G.P. Salam, {\em Nucl. Phys} {\bf B 511} (1998) 396.
%\bibitem{paper:resummedc}S. Catani, B.R. Webber, {\em Phys. Lett.} {\bf B 427} (1998) 377.
\bibitem{paper:h1}H1 Collaboration, C. Adloff et al., accepted by {\em Eur. Phys. J.} {\bf C}, hep-ex/9912052.
%\bibitem { zeus reference? }
\bibitem{paper:disent}S. Catani and M.H. Seymour, {\em Nucl. Phys} {\bf B 485} (1997) 291.
\bibitem{paper:newmilan}M. Dasgupta et al., {\bf JHEP 11} (1999) 25, Yu.L. Dokshitzer, {\bf JHEP 10} (1998) 1.
\bibitem{paper:revisit}Yu.L. Dokshitzer, G. Marchesini, G.P. Salam, {\em Eur. Phys. J.} {\bf C3} (1999) 1.
\end{thebibliography}
\end{document}